\documentclass[amsmath,amssymb]{revtex4}
\usepackage[dvips]{graphicx}
\usepackage{dcolumn}
\usepackage{bm}

\newcommand{\beq}{\begin{equation}}
\newcommand{\eeq}{\end{equation}}
\newcommand{\bea}{\begin{eqnarray}}
\newcommand{\eea}{\end{eqnarray}}

\begin{document}
\title[]
{Spin chiral current induced by curvature of space-time}

\author{Sergey N. Andrianov}
\email{adrianovsn@mail.ru} \affiliation{Scientific center for
gravity wave studies ``Dulkyn'', Kazan, Russia}

\author{Rinat A. Daishev }
\email{Rinat.Daishev@ksu.ru} \affiliation{$^1$ Department of
Mathematics and Department of Physics, Kazan Federal University,
Kremlevskaya str. 18, Kazan 420008, Russia}

\author{Sergey M. Kozyrev}
\email{Sergey@tnpko.ru} \affiliation{Scientific center for gravity
wave studies ``Dulkyn'', Kazan, Russia}

\begin{abstract}\noindent
Klein-Gordon equation is derived for a particle in the brane model of Universe. It is compared with squared Dirac-Fock-Ivanenko equation and expression for a chiral current is obtained by this comparison. This expression defines chiral current through variation of spin connection gauge field that arises due to the symmetry in respect to local Lorenz transformations. So, the second derivative of gravitational gauge field determines variation of chiral current responsible for variation of mass. The role of these processes on the early stages of Universe evolution is discussed. \\
\phantom{a}\\
PAC numbers: 04.20, 03.50
\end{abstract}

\maketitle \noindent

\section{Introduction}\noindent

The evolution of particles in curved space can be described by
generalized Klein-Gordon (KG) equation derived by Schr\"{o}dinger
\cite{Schr}. KG equation in the same form was derived within brane
model in paper \cite{Andrianov} and thus it can be valid just
after Big Bang and brane formation. But KG  equation does not
properly describe probability features of wave function. Besides,
it properly describes position of particle in space but gives the
solutions not only with positive time flaw but, also, with
negative one. Therefore, motion of particles in our nowadays
Universe with definite direction of local time flaw must be
described by Dirac equation. Its generalization in curved space is
named Dirac-Fock-Weyl (DFW) equation \cite{Fock, Weyl} or in other
notations Dirac-Fock-Ivanenko (DFI) equation
\cite{Fock,FockI,FockI29}.

Comparison of KG equation with squared DFW equation was performed
in paper \cite{Alhaidari}. But spin connection in DFW equation was
changed with that by some indefinite scalar matrix which quadrate
was also included in KG equation therein. We perform here rigorous
comparison of KG equation in its original form \cite{Schr} with
squared DFI equation and obtain formula for chiral current.
Finally, we discuss generation of mass connected with chiral
current variation on early stages of Universe evolution.


\section{Squared Dirac-Fok-Ivanenko equation}
We have derived KG equation for a brane that has the following
form \cite{Andrianov}

\begin{equation}
g^{i j}\nabla_i \nabla_j \psi =\left\{ \frac{1}{4} R
-\left(\frac{m c}{\hbar}\right)^2 \right\}\psi \label{eq1}
\end{equation}
coinciding with that of \cite{Schr}, where $\nabla_i$ is covariant
derivative, $R$  is curvature, $m$  is mass of a particle. Let's
consider, also, DFI equation
\begin{eqnarray}\label{eq2}
i \gamma^i(\nabla_i +\Gamma_i) \psi = m \psi
\end{eqnarray}
with Dirac gamma matrixes $\gamma^i$  and spin connection
$\Gamma_i$. Squaring of this equation yields
\begin{eqnarray}\label{eq3}
\left(\gamma^i \gamma^j \nabla_i \nabla_j + \left(
\gamma^i(\nabla_i \gamma^j)+\gamma^i \{\Gamma_i,
\gamma^j\}\right)\nabla_j + \gamma^i \nabla_i(\gamma^j \Gamma_j) +
(\gamma^i \Gamma_i)(\gamma^j \Gamma_j)\right)\psi  = - m^2 \psi.
\end{eqnarray}
If
\begin{equation}
\frac{1}{2}\{\gamma^i ,\gamma^j\}= g^{i j},  \label{eq4}
\end{equation}
\begin{equation}
\gamma^i(\partial_i\gamma^j)+\{(\gamma^i\Gamma_i),\gamma^j \}= 0,
\label{eq5}
\end{equation}
and
\begin{eqnarray}
 \gamma^i (\partial_i \gamma^j)B_j
+\gamma^i \partial_i(\gamma^j \Gamma_j) + (\gamma^i
\Gamma_i)(\gamma^j \Gamma_j) + (\gamma^i \Gamma_i)(\gamma^j B_j) +
(\gamma^i B_i)(\gamma^j \Gamma_j)
 = -\frac{1}{4}R, \label{eq6}
\end{eqnarray}

we come to Klein-Gordon equation (\ref{eq1})

Relations (\ref{eq4}, \ref{eq5}) coincide with that of
\cite{Alhaidari}. Taking into account expression from
\cite{Weldon}
\begin{eqnarray}
\partial_i \gamma^j + [\Gamma_i,\gamma^j]+\Gamma^j_{ik}\gamma^k=0, \label{eq7}
\end{eqnarray}
we come as in \cite{Alhaidari} from (\ref{eq4}) and (\ref{eq5}) to
the following relation between spin connection and Cristoffel
symbol.
\begin{equation}\label{eq8}
\Gamma_i  =\frac{1}{2} g^{i k} \Gamma^j_{i k}.
\end{equation}
Let's consider relation  (\ref{eq6}). Using  (\ref{eq7}) again we
get
\begin{equation}
\gamma^i \gamma^j \nabla_i \Gamma_j+ \gamma^i \gamma^j \Gamma_i
\Gamma_j =-\frac{1}{4}R, \label{eq9}
\end{equation}
where $\nabla_i \Gamma_j = \partial_i \Gamma_j - \Gamma^j_{i k}
\Gamma_k$.  Therefore, we obtain that

\begin{equation}
\gamma^i \gamma^j D_i \Gamma_j =-\frac{1}{4}R. \label{eq10}
\end{equation}
where $D_i = \nabla_i + \Gamma_i$  is generalized covariant
derivative.

\section{Chiral current.}\noindent

Let's consider the first term in (\ref{eq10}). It can be
decomposed into commutator and anticomutator parts
\begin{equation}\label{eq11}
  \gamma^i \gamma^j D_i \Gamma_j =\frac{1}{2}\{\gamma^i ,\gamma^j\}D_i \Gamma_j +
  \frac{1}{2}[\gamma^i ,\gamma^j]\left(\frac{1}{2}(D_i \Gamma_j + D_j \Gamma_i)+\frac{1}{2}(D_i \Gamma_j - D_j \Gamma_i) \right).
\end{equation}

It can be rewritten as
\begin{equation}\label{eq12}
  \gamma^i \gamma^j D_i \Gamma_j =g^{ij} D_i \Gamma_j +
  \frac{1}{4}[\gamma^i ,\gamma^j]\left(\nabla_i \Gamma_j - \nabla_j \Gamma_i+\Gamma_i \Gamma_j - \Gamma_j \Gamma_i \right).
\end{equation}

Introducing spin curvature
\begin{equation}\label{eq13}
\Phi_{i j} =\nabla_i \Gamma_j - \nabla_j \Gamma_i+\Gamma_i
\Gamma_j - \Gamma_j \Gamma_i = D_i\Gamma_j -D_j\Gamma_i
\end{equation}
we get from (\ref{eq12}) the expression
\begin{eqnarray}\label{eq14}
[\gamma^i ,\gamma^j]\Phi_{ij} &=&  -4 g^{ij} D_i \Gamma_j - R.
\end{eqnarray}
that can be used for determination of $[\gamma^i ,\gamma^j]$.

Another variant of this formula is derived as following. Spin
curvature can be expressed as \cite{Weldon}

\begin{eqnarray}
\Phi_{i j} &=&-\frac{1}{8}  [\gamma^l ,\gamma^k] R_{ijlk}.
\label{eq15}
\end{eqnarray}

Substitution of (\ref{eq15}) into (\ref{eq14}) gives
\begin{equation}
\left(\frac{1}{8}[\gamma^i ,\gamma^j][\gamma^\alpha
,\gamma^\beta]- \frac{1}{4}\{\gamma^j ,\gamma^\beta\}\{\gamma^i
,\gamma^\alpha\}\right)R_{ij\alpha\beta}=4 g^{ij} D_i \Gamma_j,
\label{eq16}
\end{equation}
Using symmetry properties of metric tensor and Bianki identity we
get
\begin{equation}
\gamma^i \gamma^j \gamma^k \gamma^l R_{i l j k} = 4 g^{i j} D_i
\Gamma_j, \label{eq17}
\end{equation}
The left hand side of (\ref{eq17}) can be rewritten as
\begin{equation}
\gamma^i \gamma^j \gamma^k \gamma^l R_{i l j k} = \delta^{i j p
s}_{k l m n} \gamma^k \gamma^l \gamma^m \gamma^n R_{is j p},
\label{eq18}
\end{equation}

where tensor $\delta^{i j p s}_{k l m n}$ is the generalized
Kronecker symbol. Further, we use the identity
\begin{equation}
\delta^{i j p s}_{k l m n} =\frac{1}{4!}\varepsilon^{i j p
s}\varepsilon_{k l m n}. \label{eq19}
\end{equation}
substituting it in (\ref{eq18}) to get
\begin{equation}
\gamma^i \gamma^j \gamma^k \gamma^l R_{i l j k} = -i \gamma^5
\widetilde{R}, \label{eq20}
\end{equation}
where
\begin{equation}
\gamma^5 =\frac{i}{4!}\varepsilon_{k l m n}\gamma^k \gamma^l
\gamma^m \gamma^n, \label{eq21}
\end{equation}
and
\begin{equation}
\widetilde{R} = \varepsilon^{i j p s} R_{i j p s}. \label{eq22}
\end{equation}
Eventually, we get from (\ref{eq17}) and (\ref{eq20}) the
following expression for $\gamma^5$:
\begin{equation}
\gamma^5  = 4i\widetilde{R}^{-1} g^{i j} D_i \Gamma_j \label{eq23}
\end{equation}
or
\begin{equation}
\gamma^5  = 4i\widetilde{R}^{-1} D_i \Gamma^i.\label{eq24}
\end{equation}

 $\gamma^5$ can be used for the determination of chiral spin current according to the formula
\begin{equation}
\gamma^{k 5}  = \overline{\psi}\gamma^{k}\gamma^{ 5}
\psi,\label{eq25}
\end{equation}
 where $\psi$  is spinor wave function. Substitution of (\ref{eq24}) into (\ref{eq25}) yields
\begin{equation}
j^{k 5}  = 4i\overline{\psi}\gamma^{k}\widetilde{R}^{-1} D_i
\Gamma^i \psi.\label{eq26}
\end{equation}

Spin connection in equation (\ref{eq26}) can be regarded as gauge
gravitational field related to the symmetry of local Lorenz
transformations. $D_i \Gamma^i $  in (\ref{eq26}) is gravitational
force. When the brane radius exceeds critical value gravitational
force supporting the quark-gluon \cite{Risi, Saaidi} brane becomes
weaker than the influence of strong interaction and quarks combine
into separate elementary particles. Variation of chiral current
describes generation of mass when quarks combine producing the
particles. So, variation of our chiral current is connected with
creation and decay of elementary particles on the background of
strongly varying gravitational field in the early Universe.


\section{Conclusion}\noindent

Thus, we have considered Klein-Gordon equation for a particle on
brane derived by us earlier using variation principle. Solutions
of Klein-Gordon equation have time reversal symmetry and thus
brane after its emergence due Big Bang supported this symmetry.
But then this symmetry was spontaneously broken and we must deal
now with Dirac equation. However, time reversing symmetry breaking
still does not yield definite time arrow.  The result of this
symmetry breaking is the emergence of particlå's spin that can be
considered as rotation of particle in extra dimension. Particle
can not move already in additional dimension because of space
compactification but it can rotate in it.  Massive particle
rotating in extra dimension produces gravitational field on brane
and thus spin connectivity through this field.

Spin projection on momentum direction defines helicity of the
particle that coincides with chirality for massless particles. No
coordinate transformation can change the helicity of massless
particle since no physical body can move faster than light. So,
particle can exist simultaneously in local times $\tau$  and
$-\tau$ for any given value of brane radius that can be regarded
as global time since these states do not mix. In the case of
massive particles, helicity coincides with chirality only when the
symmetry to parity transformations is valid for a particle. Here,
coordinate transformation can change the sign of spin projection
on particle's momentum direction but parity transformation
(inversion of coordinates) returns it back. So, particle can again
be simultaneously characterized by two contra-directional local
times.  Only chiral symmetry breaking leads eventually to
collision of local times $\tau$  and $-\tau$ and formation of time
arrow with definite time $t$. With that, inertial mass of the
particle becomes larger because now the particle does not
experience free evolution with times $\tau$  and $-\tau$ but
evolution in definite time $t$ with a process of time arrow
formation.

Squaring Dirac-Fock-Ivanenko equation gives Klein-Gordon equation.
We have derived the expression for gamma five matrix through the
derivative of spin connection by the comparison of squared DFI
equation with KG equation and used it for determination of chiral
spin current. This chiral spin current is anomalous spin current
corresponding to spontaneous chiral symmetry breaking of mass
particle in the space of DFI equation solutions. It describes
creation and decay of elementary particles and is connected with
variation of gravitational gauge field gradient that could be
rather intensive in early Universe. Generation of mass is one of
crucial steps in the evolution of early Universe. 99 percents of
nucleon mass comes out of light quarks combination in more heavy
hadrons as a result of this chiral symmetry breaking.

\end{document}